\documentclass[12pt]{article}

\usepackage[a4paper,left=2cm,right=2cm,top=1cm,
bottom=2cm]{geometry}

\usepackage{times}
\usepackage{amsmath}
\usepackage{siunitx}
\usepackage[T1]{fontenc}
\usepackage[english]{babel}
\usepackage[latin2]{inputenc}  
\usepackage{epsfig}
\usepackage{epstopdf}
\usepackage{multirow}
\usepackage{graphicx}
\usepackage{caption}
\usepackage{url}
\usepackage{float}
\usepackage{array}
\usepackage{multirow}
\usepackage{amssymb}
\usepackage{caption}
\usepackage{subcaption}
\usepackage{titlesec}
\usepackage{cite}
\usepackage[table]{xcolor}
\usepackage{adjustbox} 
\usepackage{lipsum}
\usepackage[labelfont=bf]{caption}
\usepackage{color}

\usepackage[deletedmarkup=xout]{changes}
\definechangesauthor[color=blue]{SW}
\definechangesauthor[color=red]{TL}
\definechangesauthor[color=orange]{VM}
\definechangesauthor[color=magenta]{ZD}

\makeatletter
\def\thickhline{%
  \noalign{\ifnum0=`}\fi\hrule \@height \thickarrayrulewidth \futurelet
   \reserved@a\@xthickhline}
\def\@xthickhline{\ifx\reserved@a\thickhline
               \vskip\doublerulesep
               \vskip-\thickarrayrulewidth
             \fi
      \ifnum0=`{\fi}}
\makeatother

\newenvironment{sciabstract}{%
\begin{quote} }
{\end{quote}}

\newcounter{lastnote}

\title{Electron power absorption in micro atmospheric pressure plasma jets driven by tailored voltage waveforms in He/N$_2$}

\author
{M\'at\'e Vass$^{1,2\ast}$,  Sebastian Wilczek$^{1}$, Julian Schulze$^{1,3}$, Zolt\'an Donk\'o$^{2}$, \\
}

\date{}

\begin{document}

\maketitle 
 \vspace{-1cm}
 {\small
\begin{flushleft}

 $^{1}$  Department of Electrical Engineering and Information Science, Ruhr-University Bochum, D-44780, Bochum, Germany \\
 
 $^{2}$  Institute for Solid State Physics and Optics, Wigner Research Centre for Physics, H-1121 Budapest, Konkoly-Thege Mikl\'os str. 29-33, Hungary\\

$^{3}$ Key Laboratory of Materials Modification by Laser, Ion, and Electron Beams (Ministry of Education), School of Physics, Dalian University of Technology, Dalian 116024, People's Republic of China \\

\vspace{0.5cm}
E-mail: vass@aept.ruhr-uni-bochum.de

\end{flushleft}
}
\vspace{0.3cm}

\begin{sciabstract}
In atmospheric pressure capacitively coupled microplasma jets, Voltage Waveform Tailoring (VWT) was demonstrated to provide ultimate control of the Electron Energy Distribution Function (EEDF), which allows to enhance and adjust the generation of selected neutral species by controlling the electron power absorption dynamics. However, at the fundamental level, the physical origin of these effects of VWT remained unclear. Therefore, in this work, the electron power absorption dynamics is investigated in a He/N$_2$ jet with a nitrogen concentration of 0.05\% driven by a valleys voltage waveform at a base frequency of 13.56 MHz for different numbers of harmonics using a self-consistent Particle in Cell simulation coupled with a spatio-temporally resolved analysis of the electron power absorption based on moments of the Boltzmann equation. Due to the local nature of the transport at atmospheric pressure, ohmic power absorption is  dominant. Increasing the number of harmonics, due to the peculiar shape of the excitation waveform the sheath collapse at the grounded electrode is shortened relative to the one at the powered electrode. As a consequence, and in order to ensure flux compensation of electrons and positive ions at this electrode, a high current is driven through the discharge at the time of this short sheath collapse. This current is primarily driven by a high ohmic electric field. Close to the grounded electrode, where the electron density is low and the electric field is, thus, high, electrons are accelerated to high energies and strong ionization as well as the formation of a local electron density maximum are observed. This electron density maximum leads to a local ambipolar electric field that acts as an electric field reversal and accelerates electrons to even higher energies. These effects are understood in detail to fundamentally explain the unique potential of VWT for EEDF control in such plasmas. 

\end{sciabstract}

\section{Introduction}

Radio frequency (RF-) driven atmospheric pressure microplasma jets ($\mu$APPJs) have gained considerable scientific attention in the last few years due to their effectiveness in generating reactive species \cite{Adamovich,r1,r2,r3,r4,r5}. This ability makes microplasma jets a versatile tool having applications in a variety of areas, such as plasma medicine (sterilization, wound healing, cancer treatment) \cite{Laroussi, Laroussi2, Graves, Weltmann,Kim, Bernhardt }, and materials processing \cite{m1,m2,m3}.  For such applications, optimizing and controlling the generation of radicals is of utmost importance. This can be achieved by controlling the electron energy distribution function (EEDF), as radicals are generated by electron impact excitation/dissociation of the background gas. This control might be realized if a thorough understanding of electron power absorption, i.e. the way electrons gain and lose their energy within the plasma, is achieved. One might be tempted to conclude that due to the high collisionality of the system, kinetic effects play a less prominent role as compared to plasmas at low pressure. However, previous investigations have shown that kinetic simulations are still preferable \cite{Iza,Eremin}. Previous investigations on electron dynamics in atmospheric pressure plasmas were conducted using fluid models \cite{fluid1, fluid2, fluid3}, hybrid models, where collisions were incorporated in a Monte Carlo Collision (MCC) approach \cite{hybrid1} and fully kinetic Particle in Cell (PIC)/MCC simulations \cite{Bischoff2018, Korolov2020, Korolov2019}.

Previous studies of electron power absorption in atmospheric pressure plasma jets ($\mu$APPJ) uncovered different power absorption mechanisms, viz. the $\Omega$- and Penning-mode \cite{om1, om2, Hemke2013, pen1, pen2} and mode transitions between them \cite{hybrid1}. In the $\Omega$-mode, electrons are mainly  accelerated at the times of sheath expansion and collapse within the RF period. Per electron, maximum power absorption occurs at positions close to the electrodes, where the electron density is low and, thus, the resistivity is high. Under such discharge conditions, the ohmic electric field is high due to the atmospheric pressure and it can effectively increase the electrons' energy. The spatio-temporal ionization and excitation dynamics observed in this power absorption mode are similar compared to those found in the $\alpha$-mode in low pressure capacitively coupled plasmas (CCP) \cite{alpha}, but the underlying mechanism of electron power absorption is entirely different. The Penning-mode will be present, if ionization maxima are found within the sheath regions at the times of maximum sheath voltage within the RF period. In contrast to the $\gamma$-mode in low pressure CCPs \cite{gamma}, where these maxima are a result of energetic secondary electrons accelerated by the sheath electric field thereby acquiring enough energy to ionize, in the atmospheric pressure case the ionization is largely due to Penning-ionization, where energetic electrons excite the helium atoms creating metastable species in the vicinity of the electrodes that can then lead to the ionization (see sec. \ref{sec3}).

Typically, $\mu$APPJs are driven by a single driving frequency \cite{Golda}. This leads to an inherent spatio-temporal symmetry of the electron power absorption dynamics within each RF period and strongly limits its control \cite{Bischoff2018}. Such control is, however, needed to tailor the EEDF to enhance the generation of specific neutral species by electron impact driven reactions with particular electron energy thresholds. Recently Voltage Waveform Tailoring (VWT) was demonstrated to provide ultimate control of the EEDF and to allow enhancing and controlling the generation of selected neutral species such as helium metastables and atomic oxygen compared to single frequency operation \cite{Gibson2019,Korolov2019,Korolov2020,Korolov2021}. The energy efficiency of such plasma processes was demonstrated to be higher compared to classical single frequency operation \cite{Korolov2021b}, which is strongly beneficial for applications. Korolov et al. demonstrated that this is caused by a breaking and control of the spatio-temporal symmetry of the electron power absorption dynamics induced by tailored voltage waveforms \cite{Gibson2019,Korolov2019,Korolov2020,Korolov2021}. For instance, driving a $\mu$APPJ by a valleys-/peaks-waveform was found to cause strong electron impact excitation and high neutral particle densities adjacent to only one electrode. 

These unique effects, however, have not been understood fundamentally up to now, i.e. the question why valleys-/peaks- driving voltage waveforms induce such a symmetry breaking in $\mu$APPJs is not answered. In low pressure CCPs, using an analysis based on the first velocity moment equation of the Boltzmann equation aided by PIC/MCC simulations, originally applied by Surendra et al. \cite{sur}, provides a fully self-consistent description of electron power absorption. In recent years, this method, also called the Boltzmann term analysis, has been applied to various low pressure CCPs including electropositive \cite{trev,schulze18, Wilczekheating, VMOhm}, electronegative gases \cite{VMo2, prot1, prot2} and magnetized discharges \cite{mag1,mag2,mag3}, by which a considerable amount of new insight into the physics of electron power absorption has been gained. Generally and to our knowledge, no fully self-consistent description of electron power absorption has been provided for $\mu$APPJs. 

In this paper, we use the Boltzmann term analysis and apply it to a He/N$_2$ $\mu$APPJ excited by a valleys voltage waveform synthesized from different numbers of harmonics to investigate the characteristics of electron power absorption at atmospheric pressure and to provide a fundamental explanation for the more energy efficient and controllable generation of reactive and excited neutrals by VWT in such microplasmas. Based on the Boltzmann term analysis, we show that contrary to the low pressure scenario, ohmic power absorption is dominant. By increasing the number of driving harmonics, due to the peculiar nature of the excitation waveform, the duration of the sheath collapse at both electrodes become different. In order to ensure flux compensation of positive ions and electrons at the electrode, where the sheath collapse is short, within each period of the fundamental driving frequency \cite{ChargeDyn}, a large electron current must be drawn to this electrode during the local sheath collapse. This is not the case at the other electrode, where the sheath collapse is much longer and electrons have more time to get to this electrode. To drive such a high current, a strong ohmic electric field is generated at the time of the short sheath collapse throughout the plasma. Due to the decrease of the electron density and, thus, the conductivity towards the electrodes, this ohmic field is space dependent and maximum close to the electrode, where the sheath collapse is short. Consequently, electrons are accelerated to high energies at this position and time within the fundamental RF period and cause strong ionization. Jointly with an enhanced Penning ionization at the grounded electrode due to locally enhanced helium metastable densities, this ionization, in turn, leads to the formation of a local electron density maximum close to this electrode. The related electron density gradient causes an ambipolar electric field that enhances the acceleration towards this electrode even further. Ultimately, this mechanism, which happens only at one electrode for a valleys  driving voltage waveform, results in a strong electric field reversal during the short local sheath collapse, which leads to the generation of highly energetic electrons locally. This explains, at a fundamental level, the superior performance of VWT for generating and controlling radical and excited species densities in $\mu$APPJs. 


The paper is structured as follows: in section \ref{sec2}, the theoretical background of the Boltzmann term analysis is introduced. Section \ref{sec3} gives an overview of the specifications of the PIC/MCC simulation method from which all physical parameters necessary for the analysis can be obtained. In section \ref{sec4}, results are presented and discussed. Finally, in section \ref{sec5} conclusions are drawn.

\section{Theoretical background}\label{sec2}

The basis of the Boltzmann term analysis is the first velocity moment equation of the 1D Boltzmann equation (i.e. the momentum balance equation). Rearranging this equation, the total electric field can be divided into three distinct terms as $E_{\rm tot}=E_{\rm in}+E_{\nabla p}+E_{\rm Ohm}$ \cite{VMOhm}, where the terms are given by
\begin{align}\label{Eterm}
E_{\rm in}&=-\frac{m}{ne}\left[\frac{\partial}{\partial t}(nu)+\frac{\partial}{\partial x}(nu^2)\right],\nonumber \\
E_{ \nabla p}&= - \frac{1}{ne} \frac{\partial}{\partial x} p_{\parallel}, \nonumber \\	
E_{\rm Ohm}&=-\frac{\Pi_{\rm c}}{ne}. \nonumber\\
\end{align} 

Here $n$, $u$, $m$ and $e$ are the density, mean velocity, mass and charge of the electrons, respectively. $p_{\parallel}$ stands for the diagonal element of the electron pressure tensor, where `parallel' refers to the direction of the electric field, which is perpendicular to the electrode surfaces. $\Pi_{\rm c}$ is the electron momentum loss due to collisions. 

\noindent
Each term in equation (\ref{Eterm}) represents a distinct physical mechanism. The inertia term, $E_{\rm in}$, is the electric field needed to balance the changes in the electron momentum. $E_{\rm Ohm}$, the ohmic electric field, is a result of collisions between electrons and the particles of the background gas. $E_{\nabla p}$, which is the electric field originating from the electron pressure gradient, can be split into two separate terms ($E_{\nabla p}=E_{\nabla n}+E_{\nabla T}$), given by
\begin{align}\label{Egradp}
E_{\nabla n}&=-\frac{T_{\parallel}}{ne}\frac{\partial n}{\partial x}, \nonumber \\
E_{\nabla T}&=-\frac{1}{e}\frac{\partial T_{\parallel}}{\partial x}.
\end{align} 
Here $T_\parallel$ denotes the parallel electron temperature (in units of eV), which is defined by, $T_\parallel = p_\parallel/n$. $E_{\nabla n}$ is, in quasineutral regions, identical to the classical ambipolar electric field \cite{schulze15}, and $E_{\nabla T}$ originates from the gradient of the parallel electron temperature. 

\noindent
Based on the above, the total electric field and the electron power absorption can be determined using input parameters taken from simulations. To obtain the power dissipated to electrons, the electric field must be multiplied by the electron conduction current density, $j_{\rm e}$. This Boltzmann term analysis allows to understand the fundamental mechanisms of electron power absorption and electric field generation by splitting the corresponding quantities up into the different terms mentioned above and by analyzing their respective spatio-temporal dynamics. For a more detailed description of the Boltzmann term analysis, we refer to \cite{schulze18,Wilczekheating}.

\section{Computational method}\label{sec3}

The simulations are based on a 1d3v
Particle-in-Cell Monte Carlo Collisions (PIC/MCC) code  \cite{PIC1,PIC2,PIC3}. The code has previously been applied to atmospheric pressure plasma jets \cite{Gibson2019, Bischoff2018, Korolov2020} and atmospheric-pressure nanosecond pulsed discharges \cite{donko-nsec}. As the implementation details of the discharge model have already been discussed in the references above, here we only give a brief overview of its main features.

Due to the complex nature of the plasma chemistry of the reactive He-N$_2$ plasma in the jet, the model is not able to account for all possible reactions. However, a simplified version of the plasma chemistry model with a limited number of species and reactions was found to reproduce the main features of the electron dynamics and the corresponding ionization and excitation profiles in the plasma \cite{Gibson2019, Bischoff2018, Korolov2020}. 

The species traced in the simulation are electrons, He$^+$, He$_2^+$, and N$_2^+$ ions. For electrons, the possible reactions are electron impact collisions (cross sections are taken from \cite{he-cs} for e$^-$+ He collisions and from \cite{n2-cs} for e$^-$+ N$_2$ collisions, where the latter is based on the Siglo cross section set, accessible via LxCat \cite{siglo}). These collisions are assumed to be isotropic, therefore, the elastic momentum cross sections are used. In case of electron impact excitation of He, we assume that 50\% of these events lead to the formation of either a singlet (2$^1$S) or a triplet (2$^3$S) metastable state by direct excitation to these levels or via cascades from higher states \cite{donko-nsec,Korolov2019}. {Unless otherwise indicated, in the rest of the paper we do not distinguish between these two types and by `metastable state' we generally mean both the singlet and triplet state.} The model distinguishes between two types of electron impact ionization:
\begin{eqnarray}
\label{eq:pr1}
{\rm e}^- + {\rm He}  &\rightarrow {\rm e}^- + {\rm e}^- + {\rm He}^+, \\
\label{eq:pr2}
{\rm e}^- + {\rm N}_2  &\rightarrow {\rm e}^- + {\rm e}^- + {\rm N}_2^+.
\end{eqnarray}

Since our studies are limited to small concentrations of nitrogen in the background gas ($c_{\rm N_2}=0.05\%$), elastic collisions are only considered between positive ions and He atoms as targets. For He$^+$ + He collisions, an isotropic and a backward scattering channel are taken into account \cite{Phelps}. For the collisions of He$_2^+$ and N$_2^+$ ions with He atoms, the corresponding Langevin cross sections are used. Ions are created either by electron impact ionization (processes \ref{eq:pr1}-\ref{eq:pr2}) or
via ion conversion:
\begin{equation}
{\rm He}^+ + {\rm He} + {\rm He} \rightarrow {\rm He}_2^+ + {\rm He},
\end{equation}
and through Penning ionization:
\begin{equation}\label{eq:penning}
{\rm He^*} + {\rm N}_2 \rightarrow {\rm He} (1^1{\rm S}) + {\rm N}_2^+ + {\rm e}^-.
\end{equation}

The treatment of the latter two processes is based on a Monte Carlo approach: Whenever He$^+$ and He$^\ast$ particles are created, random lifetimes are assigned to them based on the reaction rates \cite{Brok,Sakiyama} of these processes. After these lifetimes elapsed, the process is executed.

For the various species, different time steps are used based on their collision frequencies. As the collisionality of electrons is extremely high, these particles require the smallest time steps. An upper bound for this time step can be found by requiring $P = 1- \exp(-\nu \Delta t) \leq 0.1$ to hold for the collision probability. This results in $\Delta t_{\rm e} = 4.5 \times 10^{-14}$ s. For the ions, longer time steps are allowed based on their collisionalities. In our case, these values are $\Delta t_{\rm He^+} = 10 \, \Delta t_{\rm e}$ and $\Delta t_{\rm He_2^+} = \Delta t_{\rm N_2^+} = 100 \,\Delta t_{\rm e}$.

The excitation waveform used in this study is the ``valleys'' waveform, given by
\begin{equation}\label{eq:valley}
    \phi_N(t)=\phi_{\rm pp} \sum\limits_{k=1}^N\frac{2(N+k-1)}{(N+1)^2} \cos(2\pi kf_{\rm b}t+(k+1)\pi),
\end{equation}
where $N$ denotes the number of harmonics, $\phi_{\rm pp}$ the peak-to-peak voltage and $f_{\rm b}$ denotes the base frequency. In this study, we use $\phi_{\rm pp}=500$ V and $f_{\rm b}=13.56$ MHz.

The gas pressure is kept constant at $p=10^5$ Pa along with a gas temperature of $T_{\rm g}=300$ K. The electrode gap is $L=1$ mm. For the electrodes, an electron reflection probability of $e_{\rm refl}=0.5$ is assumed, together with ion-induced secondary electron emission coefficients for the respective ion species given by  $\gamma_{\rm He^+} =$ 0.2, $\gamma_{\rm He_2^+} =$ 0.12, and $\gamma_{\rm N_2^+} =$ 0.07. These specifications follow \cite{Korolov2020}. Electron emission due to metastable atoms is neglected, since only a few of these atoms are expected to reach the surfaces due to the decreased diffusion speed at high pressure \cite{Phelps1955} and due the high probability of the Penning ionization. Our simulations resemble the COST reference plasma jet, a parallel plate low temperature jet that is broadly used for fundamental laboratory studies and applications \cite{Golda}. In order to get a high spatial resolution for the power absorption terms, we use $N_{\rm g}$ = 800 grid points to resolve the gap between the electrodes. The number of superparticles in the simulations is $\sim5\cdot10^4$ and data for our analysis are collected over $100-200$ RF cycles following the convergence of the simulations. Due to the huge number of time steps per RF cycle the above data collection period provides high quality data. 

\section{Results}\label{sec4}

In the following, results are presented for the ``valleys'' waveform (eq. \ref{eq:valley}) with $\phi_{\rm pp}=500$ V and $f_{\rm b}=13.56$ MHz for different number of harmonics ranging from $N=1-4$.

\begin{figure}[H]
    \centering
    \includegraphics[width=.6\textwidth]{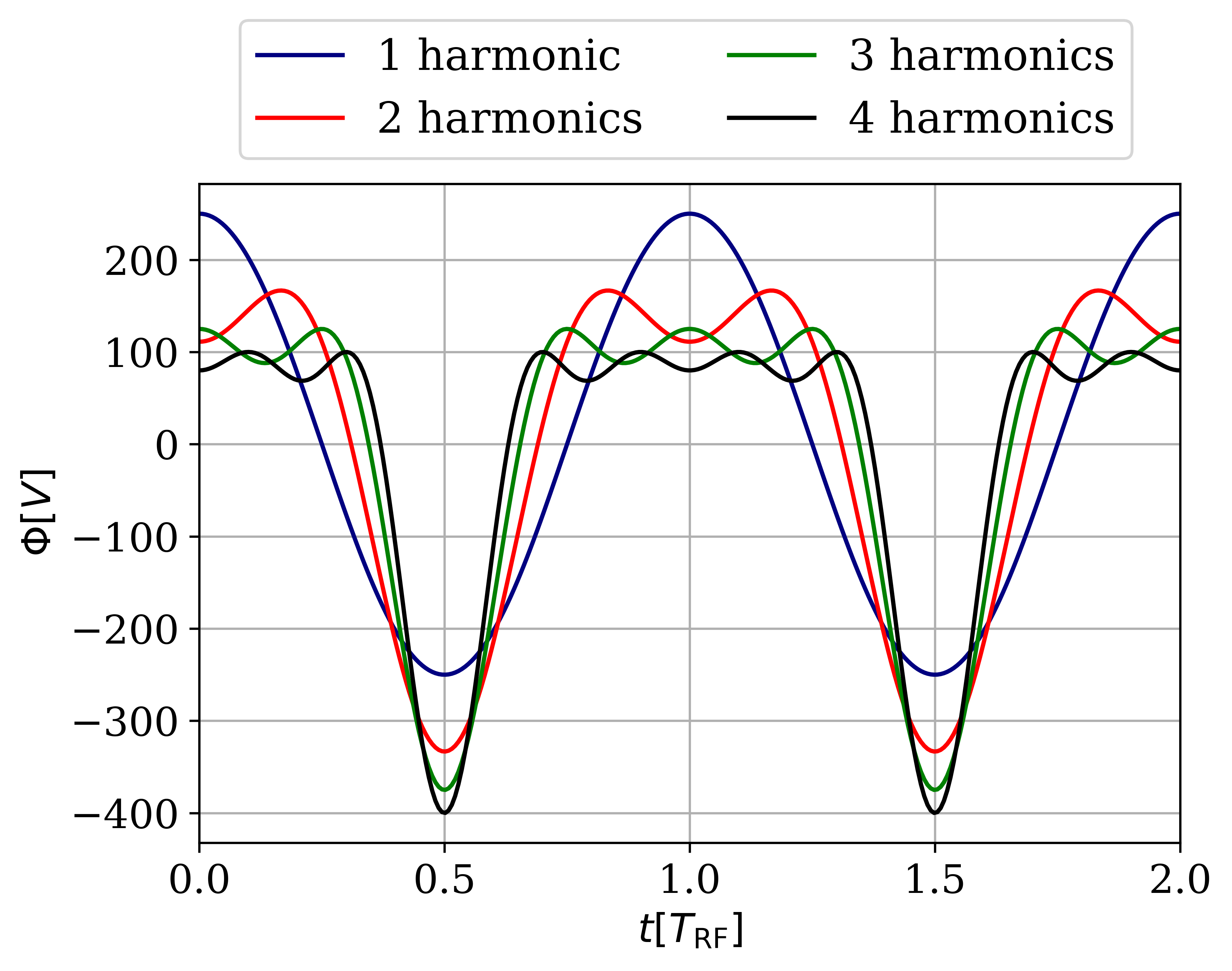}
    \caption{``Valleys'' driving voltage waveform for different numbers of harmonics.}
    \label{fig:wform}
\end{figure}

Figure \ref{fig:wform} shows the ``valleys'' driving voltage waveform given by eq. \ref{eq:valley} for different numbers of harmonics over two RF-cycles (where $T_{\rm RF}=\frac{1}{f_{\rm b}}$). The $N=1$ case is a simple cosine-shaped function with a frequency of $f_{\rm b}$. Increasing the number of harmonics has the effect of introducing a ``valley'' at $t=0.5T_{\rm RF}$, whose width decreases and depth is enlarged as the number of harmonics is increased. At the same time, the voltage is in the vicinity of its maximum for an increasing fraction of time. Thus, while in the single harmonic case the waveform monotonically decreases from its maximum towards its minimum, in case of $N=4$ harmonics, the waveform oscillates near its maximum and then sharply decreases towards its minimum. This will have very important consequences for the electron dynamics and the overall behaviour of the plasma. 

\begin{figure}[H]
    \centering
    \includegraphics[width=.6\textwidth]{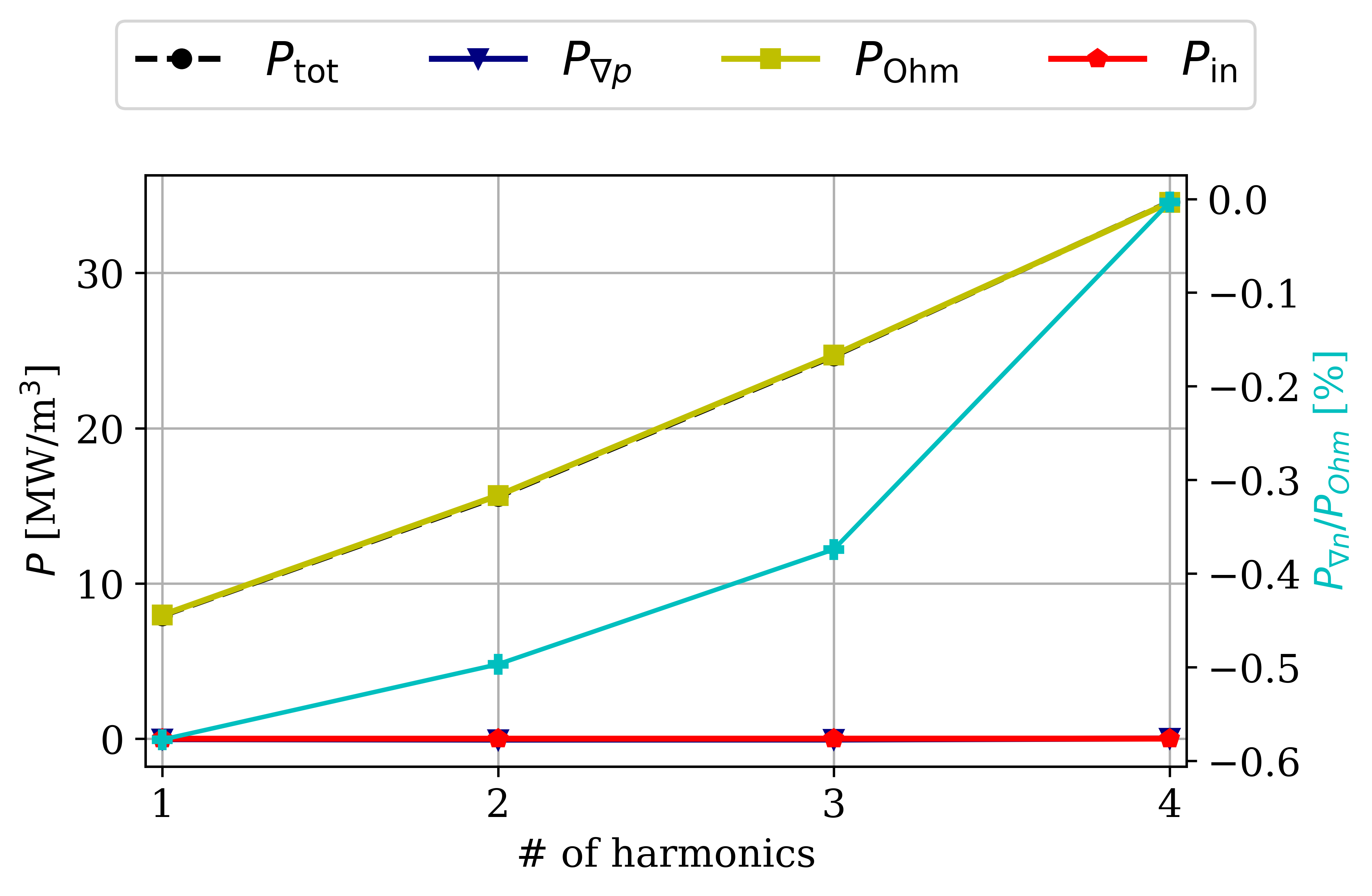}
    \caption{ Spatio-temporally averaged power absorption terms as a function of the number of harmonics as well as the proportion of ambipolar power absorption compared to the total power absorption in percentage for valleys driving voltage waveforms. Discharge conditions: $\phi_{\rm pp}=500$ V, $f_{\rm b}=13.56$ MHz, $L=1$ mm.}
    \label{fig:totpeak}
\end{figure}

To understand the properties of electron power absorption at atmospheric pressure, fig. \ref{fig:totpeak} shows the spatio-temporally averaged electron power absorption terms as well as the total electron power absorption as a function of the number of harmonics for a valleys driving voltage waveform. The first obvious observation is that ohmic heating is the dominant power absorption mechanism due to the high pressure and the corresponding high collisionality.   

\begin{figure}[H]
    \centering
    \includegraphics[width=.95\textwidth]{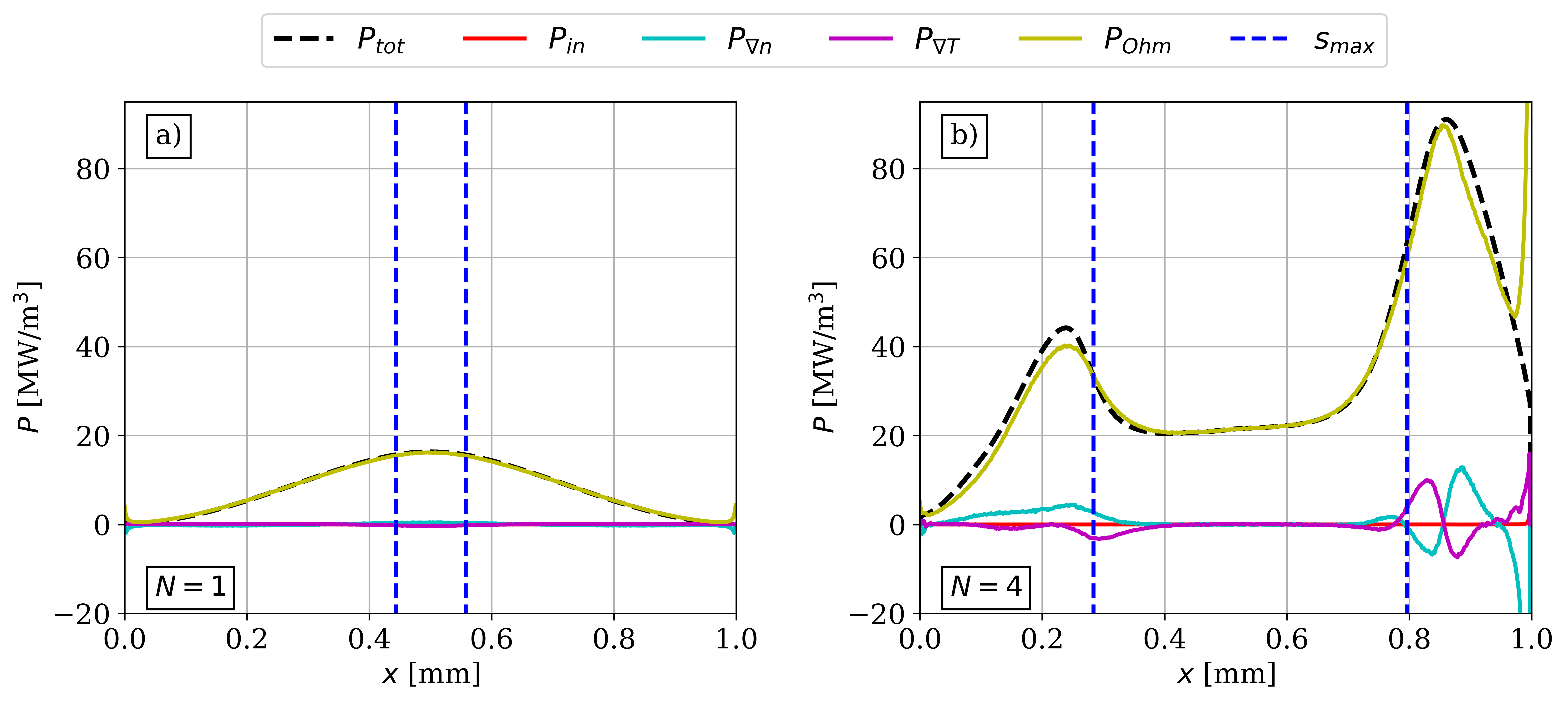}
    \caption{ Spatial distribution of the temporal average of the total electron power absorption as well as that of the different power absorption terms for valleys waveforms with 1 harmonic (a) and 4 harmonics (b). The dashed blue lines indicate the maximum of the sheath width, $s_{\rm max}$, calculated according to \cite{brsheath}. Discharge conditions: $\phi_{\rm pp}=500$ V, $f_{\rm b}=13.56$ MHz, $L=1$ mm. The powered and grounded electrodes are situated at $x=0$ and $x=1$ mm, respectively. }
    \label{fig:Ppeak}
\end{figure}

The other power absorption terms (inertial power absorption and pressure heating) are negligible on space- and time average. Even the ambipolar power absorption, which is found to be the dominant power absorption term at low pressure \cite{schulze18, Wilczekheating} is less than one percent of the total power absorption. By increasing the number of harmonics, the ohmic power absorption, and correspondingly, the total power absorption monotonically increase. In order to explain this behaviour, the details of electron power absorption need to be investigated. In the following, we will only do this in the two limiting cases, i.e. the single harmonic ($N=1$) and the four harmonics case ($N=4$).

Figure \ref{fig:Ppeak} shows the temporal average of the different power absorption terms according to eq. \ref{Eterm} along the discharge gap for a single harmonic (a) and four harmonics (b). According to panel (a), due to the atmospheric pressure, the dominant (and, essentially, the only relevant) power absorption term on time average is ohmic heating, which is a direct consequence of the atmospheric pressure and the corresponding high collisionality of electrons. This is in strong contrast to the low pressure scenario, where the dominant power absorption mechanism is usually Pressure heating \cite{schulze18, Wilczekheating}. The spatial profile of ohmic heating shows a symmetrical shape with a maximum in the middle of the discharge.  As the number of harmonics is increased, the amplitude of the different power absorption terms increases, as noted above. In this case, ohmic heating is still the dominant power absorption term, but there is a region, in the vicinity of the grounded electrode ($x=1$ mm), where $P_{\nabla n}$, the ambipolar power absorption, and $P_{\nabla T}$, the power absorption due to the temperature gradient is non-negligible. The inertial power absorption, $P_{\rm in}$, just as in the single harmonic case, is completely negligible on time average as well. Due to the difference of the excitation waveform, the shape of the temporal average of the power absorption is different compared to the single harmonic case: instead of a spatially symmetric profile with a maximum in the middle of the discharge, ohmic heating has two, unequal maxima around the maximum of the corresponding sheath widths, $s_{\rm max}$, with a plateau in the bulk region. The maximum near the grounded electrode is greater than that near the powered electrode. The maxima of $P_{\nabla n}$ and $P_{\nabla T}$ are also situated in the vicinity of the grounded electrode. At the immediate vicinity of the grounded electrode, the ohmic power absorption shows a strong increase, which is compensated by a negative ambipolar heating. The maximal sheath widths are smaller in the four harmonics case, which is an indication for the increased electron number density. To understand the physical origin of these observations regarding electron power absorption, one needs to investigate the spatio-temporal distribution of the conduction current density, $j_{\rm c}$, and given its dominance due to the high collisionality and a correspondingly local transport of the plasma, the ohmic electric field, $E_{\rm Ohm}$.

\begin{figure}[H]
    \centering
    \includegraphics[width=.9\textwidth]{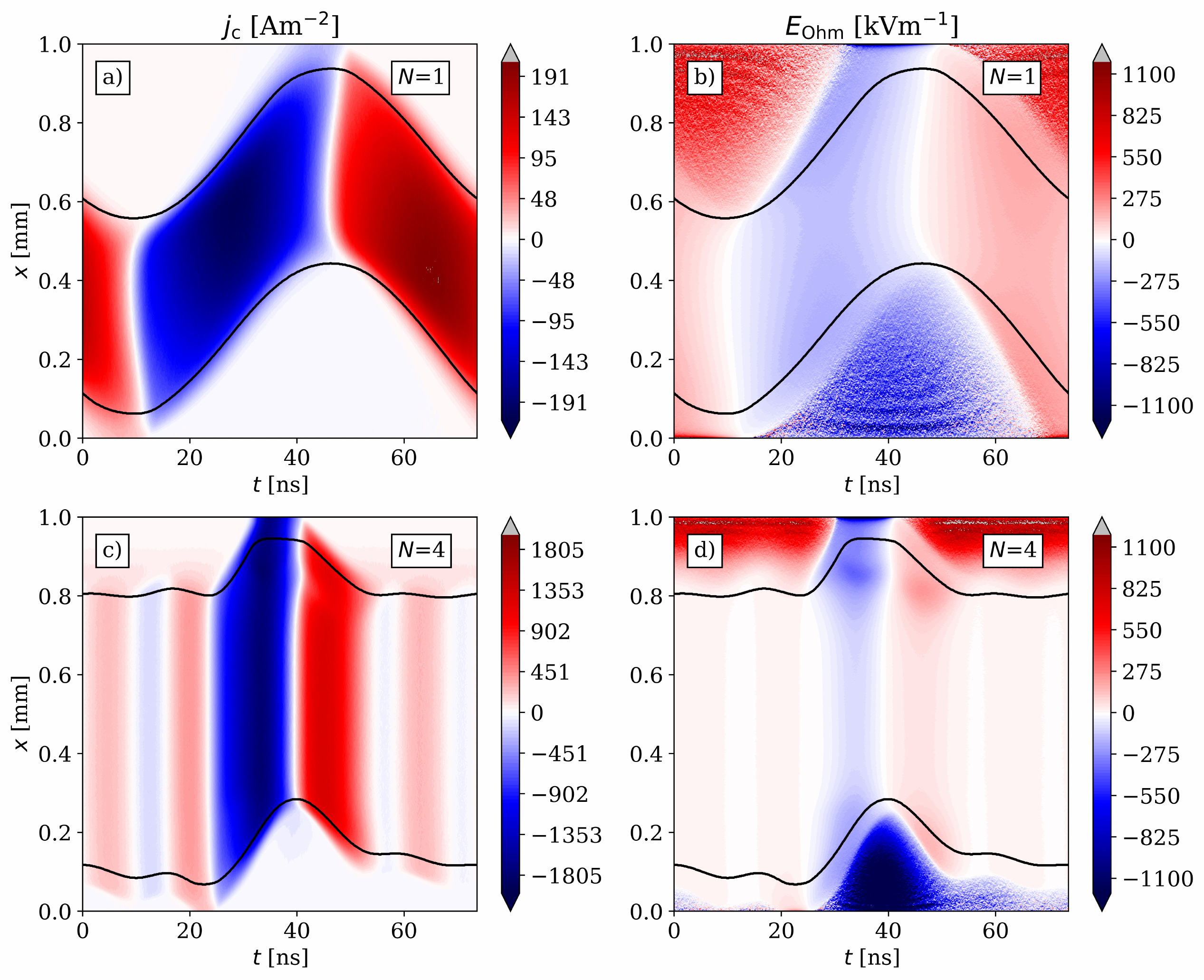}
    \caption{Spatio-temporal distribution of the electron conduction current density, $j_{\rm c}$, and ohmic electric field, $E_{\rm Ohm}$ for valleys driving voltage waveforms with 1 harmonic (a, b) and 4 harmonics (c, d). Discharge conditions: $\phi_{\rm pp}=500$ V, $f_{\rm b}=13.56$ MHz, $L=1$ mm. The black lines indicate the sheath edges. }
    \label{fig:jpeak}
\end{figure}

Figure \ref{fig:jpeak} shows the electron conduction current density, $j_{\rm c}$, (a,c) and the ohmic electric field, $E_{\rm Ohm}$, (b,d) for one and four harmonics, respectively. The current density in case of a single harmonic (a) is temporally symmetric, with the maximum during the middle of sheath expansion. Due to the local transport, it is the ohmic field that needs to drive this current through the whole discharge. Thus, in panel (b), we see a similar spatio-temporal structure for the ohmic electric field as for the current. The relatively large  maximum sheath width ($\sim0.45$ mm) explains the observed maximum in the ohmic power absorption in the discharge center in fig. \ref{fig:Ppeak}(a), as around this position the conduction current density is nonzero in the whole RF period, and given that ohmic heating is always positive, its temporal average at this position will be maximal. Changing the number of harmonics to $N=4$, a temporal asymmetry between sheath expansion and collapse is clearly visible in the conduction current density in panel (c). Due to the specific shape of the valleys driving voltage waveform, the sheath at the grounded electrode is collapsed for a short fraction of the fundamental RF period, i.e. for about 5 ns, while the sheath at the powered electrode is collapsed for a much longer time, i.e. for about 40 ns.

As the fluxes of electrons and positive ions must balance at each electrode on time average \cite{ChargeDyn} and ions flow to both electrodes continuously, a much higher electron current must flow to the grounded electrode during the short local sheath collapse compared to the situation at the powered electrode. Therefore, the electron conduction current density is higher during the sheath collapse at the grounded compared to that at the powered electrode (-1800 Am$^{-2}$ compared to 1000 Am$^{-2}$, see fig. \ref{fig:jpeak}(c)) and a higher ohmic electric field is needed to drive this higher current to the grounded electrode during the local sheath collapse (-600 kVm$^{-1}$ compared to 400 kVm$^{-1}$, see fig. \ref{fig:jpeak}(d)).  As the plasma density is lower in the vicinity of the electrodes compared to the bulk center, the ohmic electric field will increase in this region. This is the reason for the unequal maxima of the temporally averaged ohmic heating in fig. \ref{fig:Ppeak}(b). In the vicinity of the grounded electrode, two distinct local maxima can be observed in the ohmic electric field during the local sheath collapse: one in the immediate vicinity of the electrode and one on the bulk side of the sheath edge. The maximum close to the electrode surface is due to the decreased electron density at this position, and is the reason for the sharp increase in the temporally averaged ohmic power absorption around $x\approx1$ mm in fig. \ref{fig:Ppeak}(b). The other local maximum in panel (d) is also due to a local minimum of the electron density. The reason for this is as follows: Due to the low electron density in the vicinity of the grounded electrode during the local sheath collapse, there is a high instantaneous local ohmic electric field that causes strong ionization at this position. This, in turn, contributes to the formation of a local maximum of the electron density. Additionally, enhanced local Penning ionization due to a locally enlarged helium metastable density contributes to the formation of this electron density peak. The presence of this electron density maximum and the corresponding density gradient leads to the formation of an ambipolar electric field on the bulk side of this maximum that accelerates electrons towards the grounded electrode even more strongly, as $E_{\nabla n}\propto-\frac{1}{n_{\rm e}}\frac{\partial n_{\rm e}}{\partial x}$. At the position, where the electron density is maximum close to the grounded electrode, this ambipolar field as well as the ohmic field are low due to the low local density gradient and the high local electron density. Overall, this mechanism leads to the formation of a strong local maximum of $n_e$. Due to the different durations of sheath collapse at both electrodes induced by the tailored driving voltage waveform, this happens only at one and not at the other electrode. Correspondingly, electrons are only accelerated to high energies at the electrode, where the sheath collapse is short, and ionization as well as dissociation and excitation of neutrals occur predominantly at this position.

\begin{figure}[H]
    \centering
    \includegraphics[width=.95\textwidth]{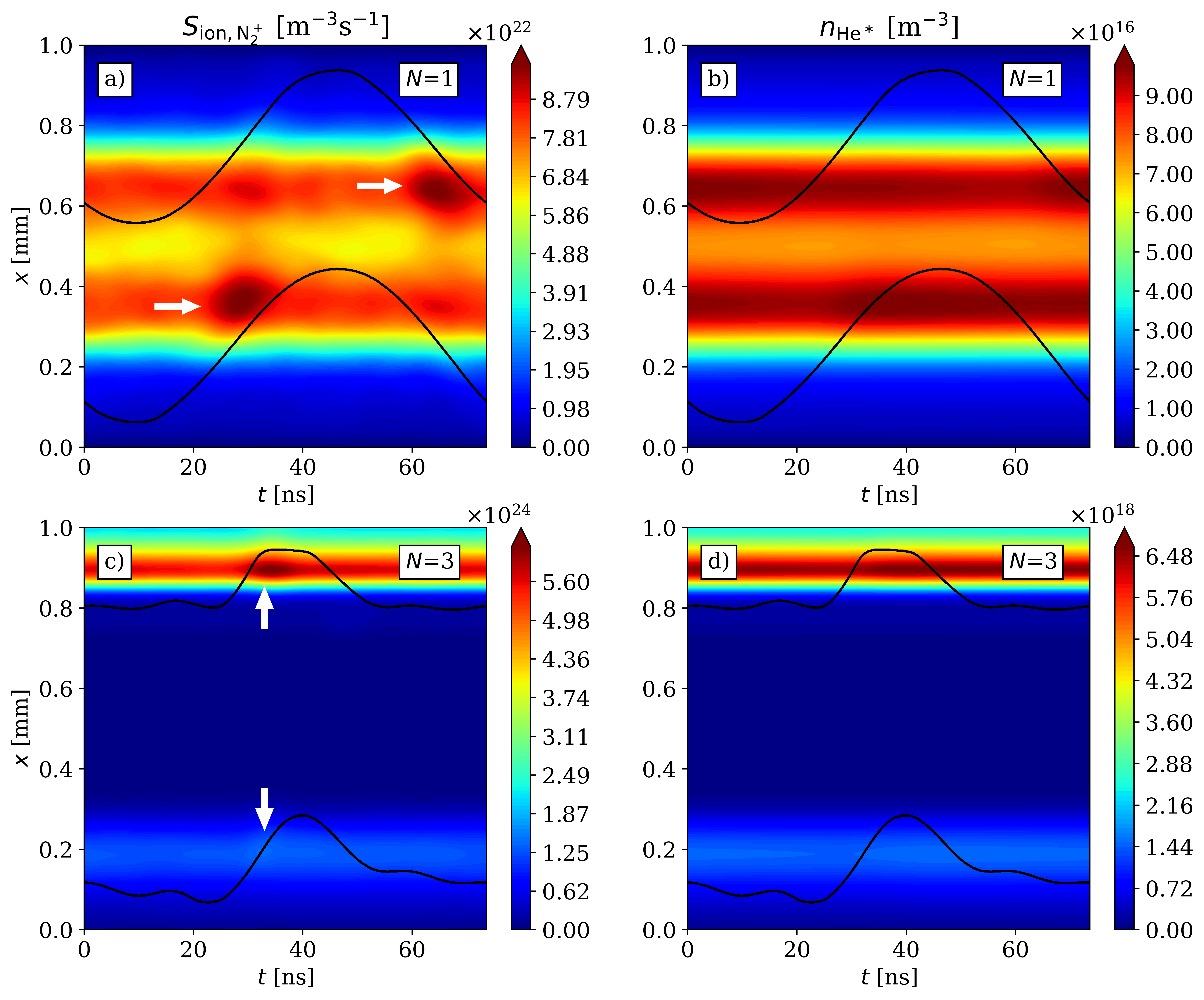}
    \caption{Spatio-temporal distribution of the ionization source function for nitrogen, $S_{\rm ion, N_2^+}$ and He metastable density, $n_{\rm He^\ast}$ for valleys driving voltage waveforms with 1 harmonic (a,b) and 4 harmonics (c,d). Discharge conditions: $\phi_{\rm pp}=500$ V, $f_{\rm b}=13.56$ MHz, $L=1$ mm. The black lines indicate the sheath edges. {The white arrows in panels (a) and (c) indicate the ionization peaks due to electron impact.}}
    \label{fig:Speak}
\end{figure}

Ultimately, this explains in detail why such tailored voltage waveforms allow to break the spatio-temporal symmetry of the electron dynamics in such discharges, which is inherently present in single frequency jets, and why the EEDF can be controlled in space and time to optimize the generation of selected radicals \cite{Bischoff2018,Korolov2019,Korolov2020,Korolov2021,Korolov2021b}. 

To see the local maximum of the ionization caused by the increased ohmic electric field during sheath collapse at the grounded electrode in case of $N=4$ harmonics, one needs to look at the possible electron impact ionization channels. As given by eqs. \ref{eq:pr1}, \ref{eq:pr2} and \ref{eq:penning}, the three possibilities are electron impact ionization of either helium atoms or nitrogen molecules, or Penning ionization. As the ionization threshold for helium is higher than that of nitrogen (24.59 eV vs. 15.6 eV), the electron impact ionization of helium can be discarded. Figure \ref{fig:Speak} shows the spatio-temporal profiles of the ionization source function for nitrogen, $S_{\rm ion, N_2^+}$ (a,c), and the He$^\ast$ metastable density, $n_{\rm He^\ast}$ (b,d) for one and four harmonics, respectively. In panel (a) the contributions from the two channels, i.e. electron impact ionization of $N_2$ and Penning ionization, can be distinguished: the local maxima during sheath expansion at either electrode are due to electron impact ionization{ (indicated by the white arrows)}, whereas the ``background'', which resembles the maxima of the helium metastable density shown in panel (b) is due to Penning ionization, as according to eq. \ref{eq:penning} Penning ionization is caused by helium metastables whose density is approximately constant in time.

Increasing the number of harmonics has the effect of introducing a spatial asymmetry to the ionization as well as the metastable density profile: the reason for this is the peculiar shape of the excitation waveform: at the grounded electrode the sheath voltage is high for most of the time within the fundamental RF-cycle. Therefore, secondary electrons can gain a high enough energy to excite the background atoms, i.e. helium, which correspondingly increases the metastable density and also the contribution of Penning ionization at this electrode. As noted above, there is a local increase of $S_{\rm ion, N_2^+}$ in panel (c), during sheath collapse at the grounded electrode, which is a consequence of the increased ohmic electric field in this region. This field is needed to drive a high current through the discharge to ensure flux conservation at the grounded electrode. This has important consequences for the spatio-temporal distribution of the electron density and, thus, for the ambipolar electric field, which fundamentally influences the electron dynamics.

\begin{figure}[H]
    \centering
    \includegraphics[width=.95\textwidth]{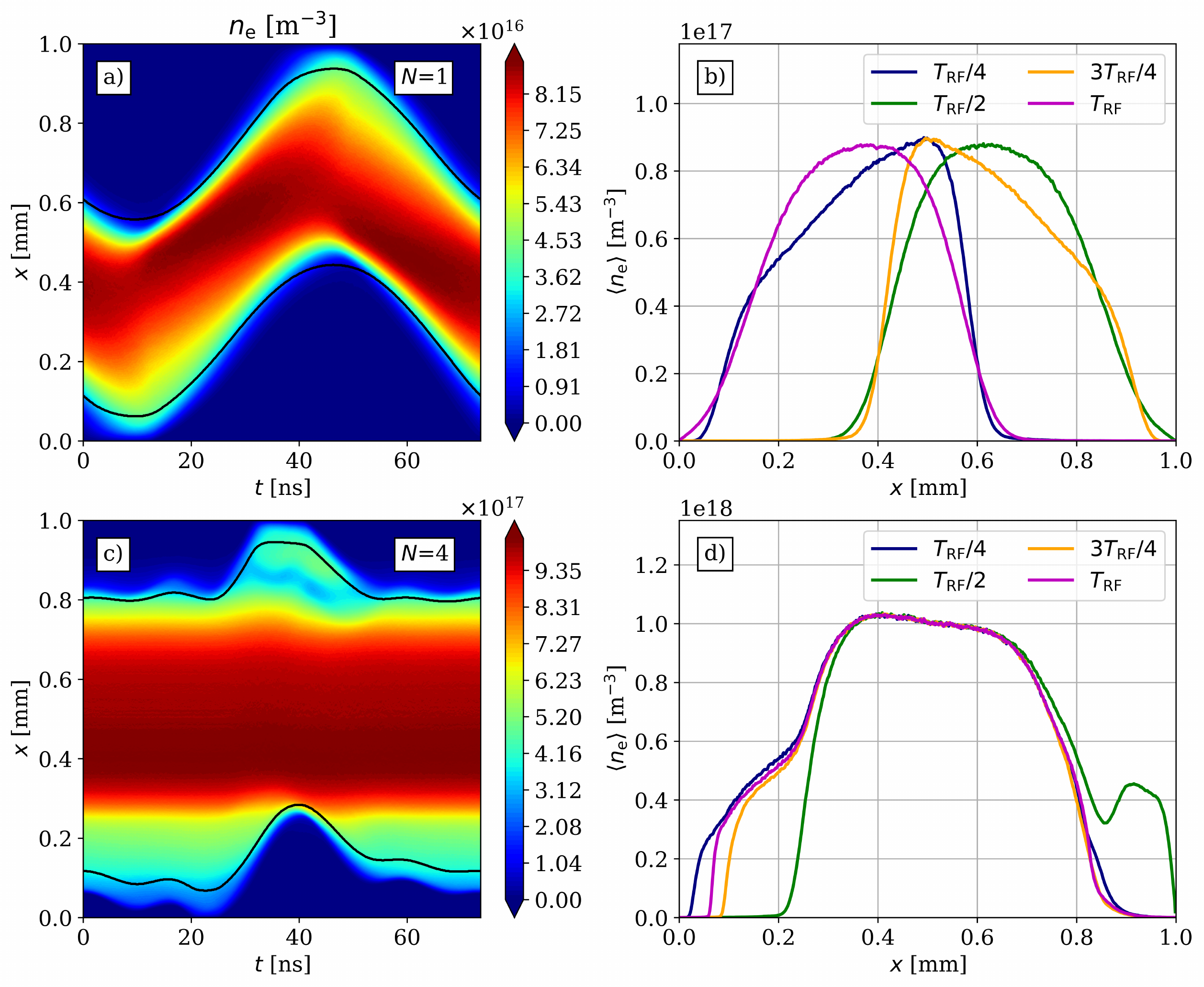}
    \caption{Spatio-temporal distribution of the electron number density, $n_{\rm e}$, and its temporal snapshots as a function of space for valleys driving voltage waveforms with 1 harmonic (a,b) and 4 harmonics (c,d). Discharge conditions: $\phi_{\rm pp}=500$ V, $f_{\rm b}=13.56$ MHz, $L=1$ mm. The black lines in panels (a,c) indicate the sheath edges.}
    \label{fig:Npeak}
\end{figure}

Figure \ref{fig:Npeak} shows the spatio-temporal profile of the electron density, $n_{\rm e}$ (a,c), and its temporal snapshots at given specific time instances in the fundamental RF-cycle as a function of position (b,d) for one and four harmonics, respectively. Panels (a) and (b) show the spatially symmetric nature of the electron density in the single frequency case: this is in accordance with the spatio-temporal profile of $j_{\rm c}$ in fig. \ref{fig:jpeak} (a). The position of the maximum of the density is temporally modulated, which is due to the relatively low electron density, which also leads to the large sheath widths. On the other hand, the electron density profile in case of four harmonics shows a completely different behaviour. 
The electron density is increased compared to the one harmonic case; the reason for this being the peculiar nature of the excitation waveform: when the number of harmonics is increased, the electrons get a stronger ``kick'' at $t=0.5T_{\rm RF}$. This increases the energy of the electrons, which can then contribute more to ionization. The waveform shape is also responsible for the slightly shifted shape of $n_{\rm e}$: the density is quite low in the vicinity of the powered electrode up to the maximum sheath width, as the sheath is relatively small during most of the time in the RF-cycle, where little ionization happens, but near half of the RF-cycle the sheath width is sharply increased resulting in higher ionization. In the vicinity of the grounded electrode during the sheath expansion/collapse phase of the ``valley'' a local electron density increase is present; this is also shown in panel (d) at the time of $t=0.5T_{\rm RF}$, near $\approx0.9$ mm the electron density first decreases and then increases as a function of position, which is a consequence of the local ionization maximum in fig. \ref{fig:Speak}(c) that is due to the local maximum of the ohmic electric field in fig. \ref{fig:jpeak}(d) and a local enhancement of the Penning ionization due to the high local metastable density. 

\begin{figure}[H]
    \centering
    \includegraphics[width=.95\textwidth]{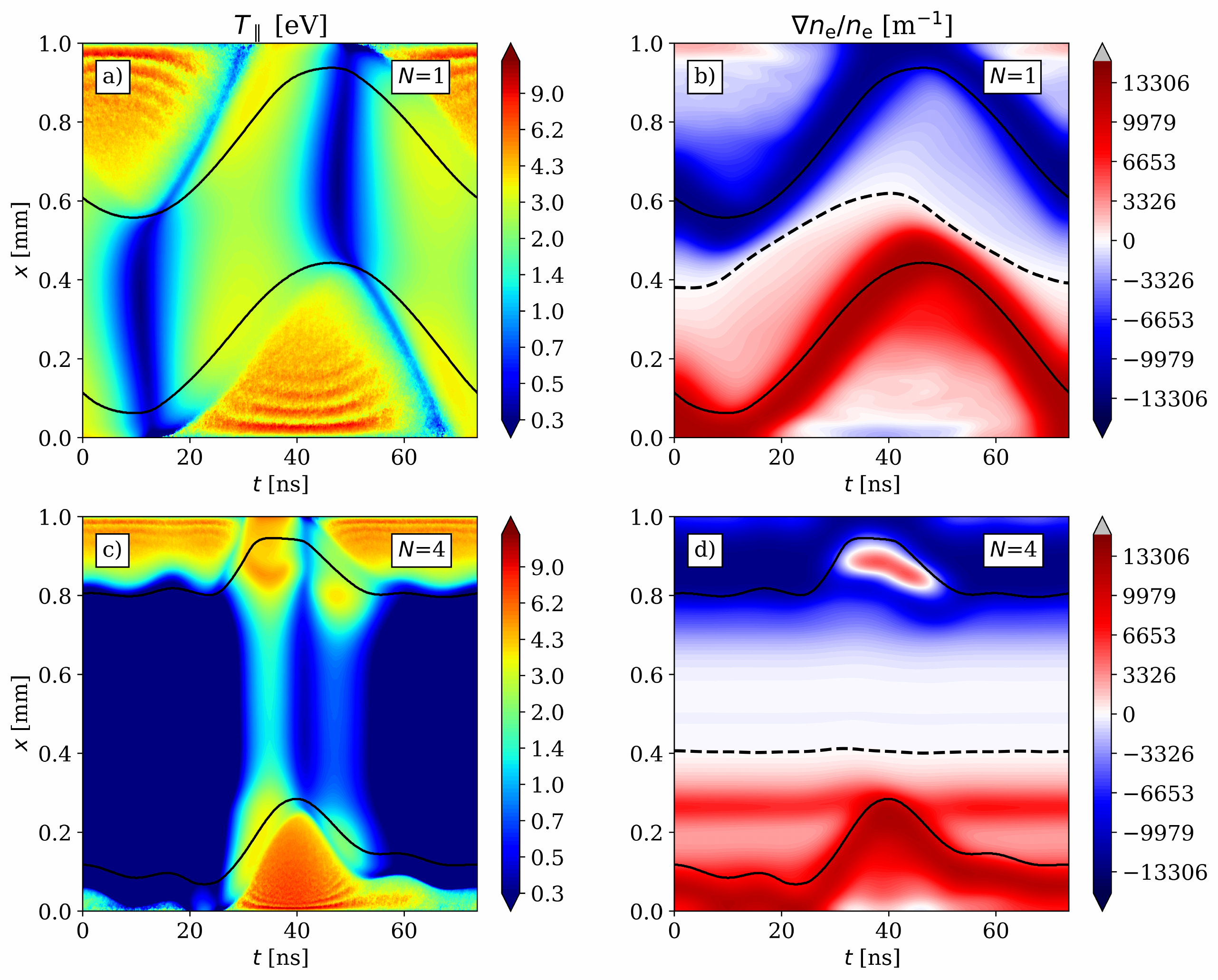}
    \caption{Spatio-temporal distribution of the parallel electron temperature, $T_{\parallel}$ and normalized electron density gradient, $\nabla n_{\rm e}/n_{\rm e}$ for valleys driving voltage waveforms with 1 harmonic (a, b) and 4 harmonics (c, d). Discharge conditions: $\phi_{\rm pp}=500$ V, $f_{\rm b}=13.56$ MHz, $L=1$ mm. The black lines indicate the sheath edges. The dashed black lines in panels (b,d) indicate the points where $\nabla n_{\rm e}$ is zero.}
    \label{fig:Tpeak}
\end{figure}

The presence of this local density increase will result in a positive density gradient between the position of the local minimum near the maximum sheath width at the grounded electrode and the position of the local maximum, resulting in an electric field reversal due to the ambipolar field.

In order to investigate the ambipolar electric field, given by $E_{\nabla n}=-\frac{T_\parallel}{n_{\rm e}}\frac{\partial n_{\rm e}}{\partial x}$, one needs to take a look at fig. \ref{fig:Tpeak}, which shows the parallel electron temperature, $T_\parallel$ (a,c) and the normalized electron density gradient, $\frac{1}{n_{\rm e}}\frac{\partial n_{\rm e}}{\partial x}$ (b,d) for one and four harmonics, respectively.

 The electron temperature in the single harmonics case (panel (a)) is increased during sheath expansion due to the fact that electrons are accelerated by the expanding sheath. At this time, within the RF period, the temperature decreases as a function of position from the sheath edge towards the bulk and then, around the position of maximum sheath edge at the opposite electrode, begins to increase again. The reason for this behaviour is the local character of the transport: as shown previously, due to the atmospheric pressure, ohmic heating is the dominant power absorption, thus, the profile of the temperature is primarily determined by the spatio-temporal distribution of the ohmic electric field. As in the bulk region the electron density, $n_{\rm e}$, is high as shown in fig. \ref{fig:Npeak}(a), thus the resistivity is low, the ohmic electric field is smaller, and as a consequence, the temperature will also decrease in this region. The spatially periodic patterns present inside the sheath are due to the small energy relaxation length of the secondary electrons, emitted from the electrodes, and are reminiscent of the Franck-Hertz experiment \cite{Korolov2020}. Panel (b) shows the normalized electron density gradient, $\frac{1}{n_{\rm e}}\frac{\partial n_{\rm e}}{\partial x}$, for the single harmonic case. As one moves away from the powered electrode, this quantity is positive, i.e. the electron density increases, then reaches its maximum (whose axial position is, under these conditions, time-dependent), and then decreases towards the grounded electrode in accordance with fig. \ref{fig:Npeak}(a).

The spatio-temporal distribution of the temperature in case of four harmonics is shown in panel (c). The temporal asymmetry between sheath expansion and collapse is present in this quantity as well, which is due to the increased current during sheath collapse at the grounded electrode to ensure flux conservation at this boundary surface. Thus, during sheath collapse at the grounded electrode, the electron temperature is higher along the whole discharge than during the collapse phase. There is a local maximum of the electron temperature at the position of the local ohmic electric field maximum in fig. \ref{fig:jpeak}(d), i.e. in the vicinity of the grounded electrode during its sheath collapse. As seen in the normalized electron density gradient, panel (d),
this is related to the local maximum of the electron density, which results in a negative electron density gradient, and, thus, a negative ambipolar field at the grounded electrode, i.e. a local electric field reversal. Thus, this electric field will increase the energy of incoming electrons accelerated by the high ohmic field in this region, resulting in more ionization and a higher local maximum of the electron density. Therefore, even though on time average the ambipolar power absorption is very small, it has an important effect on the details of the electron dynamics.

\begin{figure}[H]
    \centering
    \includegraphics[width=.95\textwidth]{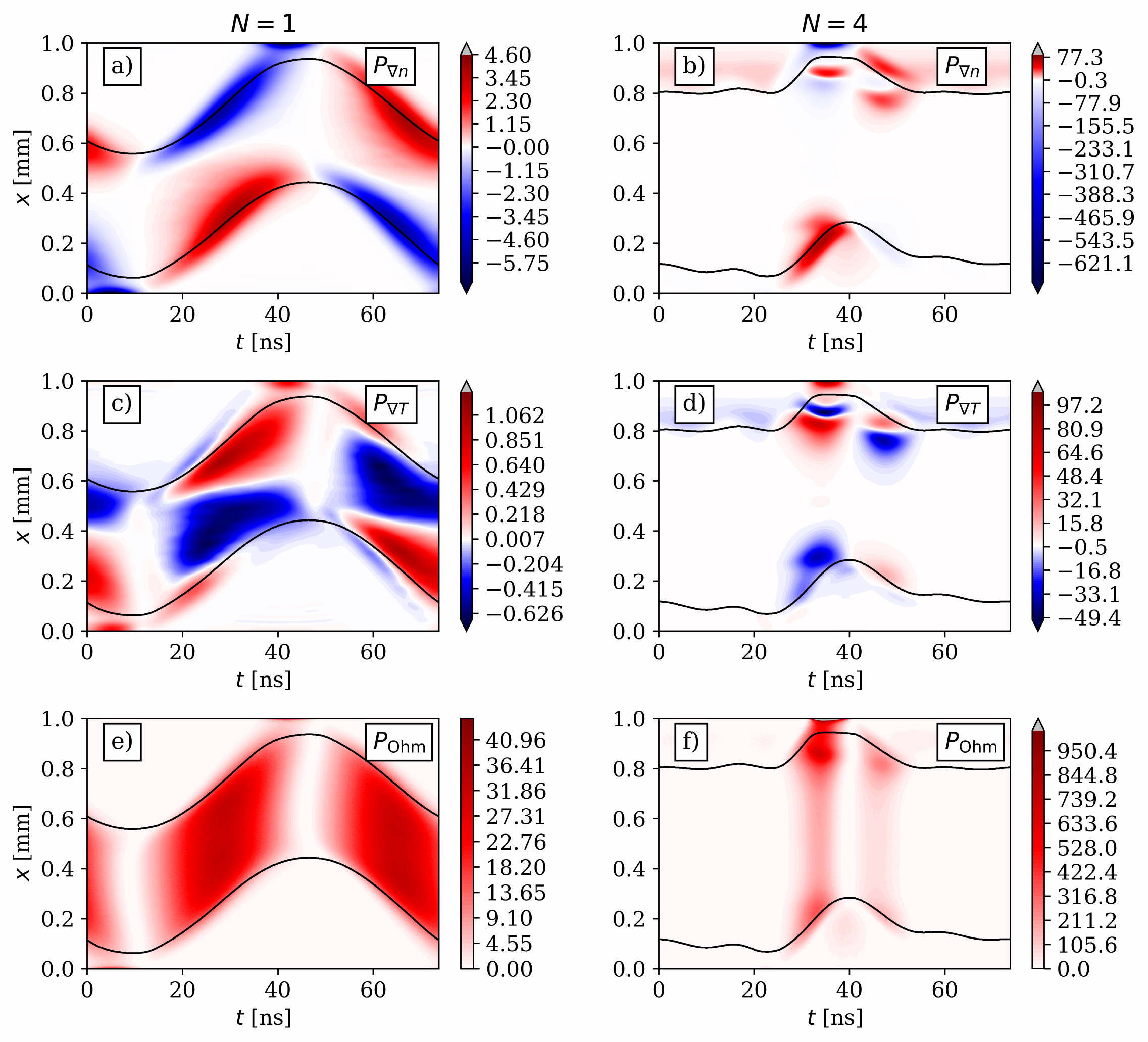}
    \caption{Spatio-temporal distribution of electron power absorption terms in units of MWm$^{-3}$ for valleys driving voltage waveforms with 1 harmonic (first column) and 4 harmonics (second column). Discharge conditions: $\phi_{\rm pp}=500$ V, $f_{\rm b}=13.56$ MHz, $L=1$ mm. The black lines indicate the sheath edges.}
    \label{fig:XT}
\end{figure}

The spatio-temporal distributions of the electron power absorption terms investigated in this paper are shown in fig. \ref{fig:XT} for one harmonic (first column) and four harmonics (second column). As the inertial power absorption proved to be negligible and hence unimportant, here we only show the other power absorption terms. 

 Panel (a) of figure \ref{fig:XT} shows the spatio-temporal distribution of the ambipolar power absorption, $P_{\nabla n}$, for a single harmonic. It is similar to that found in the case of low pressure: there is a temporally symmetric shape (although shifted {in time} due to the high pressure and the corresponding resistivity), where during sheath expansion, there is positive ambipolar heating in the vicinity of the powered electrode. Electrons are accelerated in this region by the expanding sheath, whereas during its collapse, incoming electrons are decelerated which results in a negative power absorption. In panel (c) of figure \ref{fig:XT}, the power absorption due to the temperature gradient {is negative} during sheath expansion, i.e. as the conduction current density, $j_{\rm c}$, is negative (because during sheath expansion electrons move towards the grounded electrode), $E_{\nabla T}=-\frac{1}{e}\frac{\partial T_\parallel}{\partial x}$ is positive, which indicates a negative temperature gradient in the bulk region during the first half of the RF-cycle, which then near the maximum sheath width at the grounded electrode changes sign and becomes positive. This means, that the parallel electron temperature first decreases as the electrons move away from the powered electrode and then increases as they reach the maximum sheath width of the grounded electrode, as shown in fig. \ref{fig:Tpeak}(a). Ohmic heating (panel (e) of figure \ref{fig:XT}) is positive (meaning that electrons lose momentum during collisions as they move towards the opposite electrode) in the entire discharge region. Note, that in accordance with fig \ref{fig:Ppeak} (a), the magnitude of ohmic heating is much higher than that of the other two power absorption terms.

As the number of harmonics is increased, the spatio-temporal distributions of the power absorption terms show notable differences as previously discussed. Panels (b) and (d) of figure \ref{fig:XT}, i.e. $P_{\nabla n}$ and $P_{\nabla T}$ show the presence of the electric field reversal in the vicinity of the grounded electrode. Similarly, there is a temporal asymmetry in case of ohmic heating (panel (f)), which is the result of the strong ``kick'' of the rapid sheath expansion at the grounded electrode. Due to this a high current has to flow through the discharge in order to ensure flux compensation at the grounded electrode. At positions where the electron density is low in the spatial region near the grounded electrode during the local sheath collapse, the ohmic power absorption shows a local maximum, as due to the low density the resistivity of the plasma increases. At the position of the local electron density maximum close to the grounded electrode, the ohmic power absorption is depleted. The ambipolar power absorption follows the gradient of the electron density caused by its local maximum. The behaviour of $P_{\nabla T}$ is a direct consequence of the spatio-temporal dynamics of the electron temperature discussed before.

\section{Conclusions}\label{sec5}

The electron power absorption dynamics in a micro atmospheric pressure RF plasma jet driven by a ``valleys'' waveform with 500 V peak-to-peak voltage and 13.56 MHz base frequency operated in a He-N$_2$ gas mixture with a nitrogen concentration of 0.05\% and an electrode gap of 1 mm has been investigated for one and four harmonics using the Boltzmann term analysis, which offers a self-consistent, spatio-temporally resolved description of electron power absorption \cite{schulze18, Wilczekheating, VMOhm}. The input parameters needed for the analysis were obtained by 1d3v PIC/MCC simulations. 

Due to the atmospheric pressure and the corresponding local nature of the transport, the dominant, and essentially only relevant power absorption term on space and time average is found to be ohmic heating. In the single frequency case, the spatial profile of the time-averaged ohmic heating has a symmetrical shape with a maximum in the center of the discharge. The reason for this is the relatively low electron density and the correspondingly large sheath widths, due to which the maximum of the electron density is time modulated. The spatio-temporal profiles of $P_{\nabla n}$ and $P_{\nabla T}$, i.e. the ambipolar power absorption and the power absorption corresponding to the gradient of the parallel electron temperature, are found to be similar to that found in low pressure CCPs (\cite{schulze18, Wilczekheating}), with the important difference that in this case these power absorption terms are found to be negligible on time average. $P_{\rm in}$, i.e. inertial power absorption is found to be completely negligible in the cases considered. 

Increasing the number of harmonics from 1 to 4 causes notable differences. Although, on space- and time average still ohmic heating is dominant, ambipolar power absorption is found to have a non-negligible contribution to the time-averaged electron power absorption in a spatial region near the grounded electrode. Furthermore, the spatial profile of the time-averaged electron power absorption becomes asymmetric, which is the consequence of the peculiar shape of the excitation waveform, as during sheath collapse at the grounded electrode electrons are strongly accelerated, which increases their energy contributing to an increased electron power absorption on time average. This is the reason for an increase of the electron density compared to the single frequency case at otherwise identical discharge conditions. An additional consequence of the ``valleys'' waveform is a temporal asymmetry in the electron conduction current, $j_{\rm c}$, as the amplitude of the current density is found to be higher during sheath collapse at the grounded electrode than during sheath expansion at this electrode. The reason for this temporal asymmetry is the different durations of sheath collapse at the electrodes: due to the valley driving voltage waveform the sheath collapse at the grounded electrode is shorter than at the powered electrode. Thus, in order to ensure flux compensation of positive ions and electrons at this electrode, the electron conduction current density during sheath collapse at the grounded electrode has to be higher compared to the sheath collapse at the other electrode. Due to the local nature of the transport at atmospheric pressure, it is primarily the ohmic electric field that needs to drive this current through the discharge and is, thus, temporally asymmetric as well. During the sheath collapse at the grounded electrode, electrons accelerated by this high ohmic field as well as a locally enhanced Penning ionization due to locally enhanced helium metastable densities lead to the generation of a local electron density maximum. In the spatial region between the maximum sheath width at the grounded electrode and this maximum, i.e. where the electron density gradient is positive, a negative ambipolar electric field is formed, leading to an electric field reversal that accelerates electrons towards the grounded electrode and enhances the local generation of energetic electrons in the vicinity of the grounded electrode. 

These findings explain at the fundamental level how the spatio-temporal electron power absorption dynamics work in $\mu$APPJs and why its spatio-temporal symmetry can be broken as well as controlled by Voltage Waveform Tailoring. This is the physical origin of the enhanced EEDF control and the enhanced generation of radicals as well as excited neutrals by Waveform Tailoring in such discharges. Thus, these insights might provide the basis for the optimization of applications of such microplasmas that rely on high/controlled fluxes of neutral particle generated by electron impact driven reactions with specific electron energy thresholds and dependencies.


\section*{Acknowledgements}
This work was funded by the German Research Foundation in the frame of the project, ``Electron heating in capacitive RF plasmas based on moments of the Boltzmann equation: from fundamental understanding to knowledge based process control'' (No. 428942393). Support by the DFG via SFB TR 87, project C1, SFB 1316, project A4 and by the Hungarian Office for Research, Development, and Innovation (NKFIH 134462) is gratefully acknowledged.


\begin{thebibliography}{90}

\bibitem{Adamovich} Adamovich I et al. 2017 {\it J. Phys. D: Appl. Phys.} {\bf 50} 323001

\bibitem{r1} Jiang J and Bruggeman P J 2021 {\it J. Phys. D: Appl. Phys.} {\bf 54}(15) 15LT01

\bibitem{r2} Kong D, Zhu P, He F, Han R, Yang B, Wang M, and Ouyang J 2021 {\it J. Appl. Phys.} {\bf 129}(10) 103303

\bibitem{r3} Thorben K, Christoph R, Maik F, J\"org I and Holger K 2021  {\it EPJ Tech. Instrum.} {\bf 7}(1)

\bibitem{r4} Ghimire B et al. 2021 {\it Plasma Sources Sci. Technol.} {\bf 30}(3) 035009

\bibitem{r5} Weerasinghe J et al. 2021 {\it Sol. Energy} {\bf 215} 367-374

\bibitem{Laroussi} Laroussi M 2005 {\it Plasma Process. Polym.} {\bf 2}  391

\bibitem{Laroussi2} Laroussi M, Lu X and Keidar M 2017 {\it J. Appl. Phys.} {\bf 122}(2) 020901

\bibitem{Graves} Graves D B 2014 {\it Phys. Plasmas} {\bf 21} 080901

\bibitem{Weltmann} Weltmann K D and von Woedtke T 2017 {\it Plasma Phys. Control. Fusion} {\bf 59} 014031

\bibitem{Kim} Kim S J, Chung T H, Bae S H and Leem S H 2009 {\it Plasma Process. Polym.} {\bf 6} 676

\bibitem{Bernhardt} Bernhardt T, Semmler M. L, Sch\"afer M, Bekeschus S, Emmert S and Boeckmann L 2019 {\it Oxid. Med. Cell. Longev.} 2019

\bibitem{m1} Babayan S E, Jeong J Y, Tu V J, Park J, Selwyn G S and Hicks R F 1998 {\it Plasma Sources Sci. Technol.} {\bf 7} 286

\bibitem{m2} Reuter S, Sousa J S, Stancu G D and van Helden J H 2015 {\it Plasma Sources Sci. Technol} {\bf 24} 054001

\bibitem{m3} Ichiki T, Tauro R and Horiike Y 2004 {\it J. Appl. Phys.} {\bf 95} 35

\bibitem{Iza} Iza F, Lee J K, and Kong M G 2007 {\it Phys. Rev. Lett.}  {\bf 99} 075004

\bibitem{Eremin} Eremin D, Hemke T and Mussenbrock T 2016 {\it Plasma Sources Sci. Technol.} {\bf 25} 015009

\bibitem{fluid1} Niermann B, Hemke T, Babaeva N Y, B\"oke M, Kushner M J, Mussenbrock T and Winter J 2011 {\it J. Phys. D: Appl. Phys.} {\bf 44} 485204

\bibitem{fluid2} Hemke T, Wollny A, Gebhardt M, Brinkmann R P and Mussenbrock T 2011 {\it J. Phys. D: Appl. Phys.} {\bf 44} 285206

\bibitem{fluid3} Liu Y, Korolov I, Hemke T, Bischoff L, H\"ubner G, Schulze J and Mussenbrock T 2021 {\it J. Phys. D: Appl. Phys.} {\bf 54} 275204.

\bibitem{hybrid1} Liu Y et al. 2020 {\it Plasma Sources Sci. Technol.}, Accepted manuscript

\bibitem{Bischoff2018} Bischoff L, H{\"u}bner G, Korolov I, Donk\'{o} Z, Hartmann P, Gans T, Held J, Schulz-von der Gathen V, Liu Y, Mussenbrock T and Schulze J 2018 {\it Plasma Sources Sci. Technol.} {\bf 27} 125009 

\bibitem{Korolov2020} Korolov I, Leimk{\"u}hler M, B{\"o}ke M, Donk\'{o} Z, Schulz-von der Gathen V, Bischoff L, H{\"u}bner G, Hartmann P, Gans T, Liu Y, Mussenbrock T and Schulze J 2020 {\it J. Phys. D: Appl. Phys.} {\bf 53}(18) 185201

\bibitem{Korolov2019} Korolov I, Z Donk\'{o}, G H\"{u}bner, L Bischoff, P Hartmann, T Gans, Y Liu, T Mussenbrock, and J Schulze 2019 {\it Plasma Sources Sci. Technol.} {\bf 28} 094001

\bibitem{om1} Schulz-von der Gathen V, Schaper L, Knake N, Reuter S,Niemi K, Gans T and Winter J 2008 {\it J. Phys. D: Appl. Phys.} {\bf 41} 194004

\bibitem{om2} Reuter S, Winter J, Iseni S, Peters S, Schmidt-Bleker A, D\"unnbier M, Sch\"afer J, Foest R and Weltmann K D 2012 {\it Plasma Sources Sci. Technol.} {\bf 21} 034015

\bibitem{Hemke2013} Hemke T, Eremin D, Mussenbrock T, Derzsi A, Donk\'o Z, Dittmann K, Meichsner J and Schulze J 2013 {\it Plasma Sources Sci. Technol.} {\bf 22} 015012

\bibitem{pen1} Schr\"oder D, Burhenn S, de los Arcos T and Schulz-von der Gathen V 2015 {\it J. Phys. D: Appl. Phys.} {\bf 48} 055206

\bibitem{pen2} D\"unnbier M, Becker M M, Iseni S, Bansemer R, Loffhagen D, Reuter S and Weltmann K D 2015 {\it Plasma Sources Sci. Technol.} {\bf 24} 065018

\bibitem{alpha} Schulze J, Heil B G, Luggenh\"olscher D, Brinkmann R P and Czarnetzki U 2008 {\it J. Phys. D: Appl. Phys.} {\bf 41} 195212

\bibitem{gamma} Belenguer P and Boeuf J P 1990 {\it Phys. Rev. A} {\bf 41} 4447

\bibitem{Golda} Golda J et al. 2016 {\it J. Appl. Phys.} {\bf 49} 084003

\bibitem{Gibson2019}  Gibson A R, Donk\'{o} Z, Alelyani L, Bischoff L, H{\"u}bner G, Bredin J, Doyle S, Korolov I, Niemi K, Mussenbrock T, Hartmann P, Dedrick J P, Schulze J, Gans T and O'Connell D 2019 {\it Plasma Sources Sci. Technol.} {\bf 28} 01LT01 

\bibitem{Korolov2021} Korolov I, Steuer D, Bischoff L, Huebner G, Liu Y, Schulz-von der Gathen V, Boeke M, T Mussenbrock, and J Schulze 2021 {\it J. Phys. D} {\bf 54} 125203

\bibitem{Korolov2021b} Korolov I, Donko Z, Huebner G, Liu Y, T Mussenbrock, and J Schulze 2021 {\it arXiv:2104.06635} 


\bibitem{sur} Surendra, M and Dalvie, M 1993 {\it Phys. Rev. E} {\bf 48}(5) 3914

\bibitem{trev} Lafleur T and Chabert P 2015 {\it Plasma Sources Sci. Technol.} {\bf 24}(4) 044002

\bibitem{schulze18} Schulze J, Donk\'o Z, Lafleur T, Wilczek S, Brinkmann R P 2018 {\it Plasma Sources Sci. Technol.} {\bf 27}(5) 055010

\bibitem{Wilczekheating} Wilczek S, Schulze J, Brinkmann R P, Donk\'o Z, Trieschmann J and Mussenbrock T 2020 {\it J. Appl. Phys.} {\bf 127}(18) 181101

\bibitem{VMOhm} Vass M, Wilczek S, Lafleur T, Brinkmann R P, Donk\'o Z and Schulze J 2020 {\it Plasma Sources Sci. Technol.} {\bf 29}(8) 085014

\bibitem{VMo2} Vass M, Wilczek S, Lafleur T, Brinkmann R P, Donk\'o Z and Schulze J 2020 {\it Plasma Sources Sci. Technol.} {\bf 29}(2) 025019.

\bibitem{prot1} Proto A and Gudmundsson J T 2020 {\it J. Appl. Phys.} {\bf 128}(11) 113302

\bibitem{prot2} Proto A and Gudmundsson J T 2021 ``Electron power absorption in radio frequency driven capacitively coupled chlorine discharge'' {\it Plasma Sources Sci. Technol.} sumbitted

\bibitem{mag1} Wang L, Wen D Q, Hartmann P, Donk\'o Z, Derzsi A, Wang X F, Song Y H, Wang Y N and Schulze J 2020 {\it Plasma Sources Sci. Technol.} {\bf 29}(10) 105004

\bibitem{mag2} Zheng B, Fu Y, Wang K, Schuelke T and Fan Q H 2021 {\it Plasma Sources Sci. Technol.} {\bf 30}(3) 035019

\bibitem{mag3} Zheng B, Wang K, Grotjohn T, Schuelke T and Fan Q H 2019 {\it Plasma Sources Sci. Technol.} {\bf 28}(9) 09LT03

\bibitem{ChargeDyn} Schulze J, Schuengel E, Donko Z, and Czarnetzki U 2010 {\it J. Phys. D} {\bf 43} 225201


\bibitem{schulze15} Schulze J, Donk\'o Z, Derzsi A, Korolov I and Schuengel E 2015 {\it Plasma Sources Sci. Technol.} {\bf 24} 015019
	



\bibitem{PIC1} Verboncoeur J P 2005 {\it Plasma Phys. Control. Fusion} {\bf 47} A231 

\bibitem{PIC2} Matyash K, Schneider R, Taccogna F, Hatayama A, Longo S, Capitelli M, Tskhakaya D and Bronold F X 2007 {\it Contrib. Plasma Phys.} {\bf 47} 595

\bibitem{PIC3} Donk\'o Z 2011 {\it Plasma Sources Sci. Technol.} {\bf 20} 24001



\bibitem{donko-nsec} Donk\'{o} Z, Hamaguchi S and Gans T 2018 {\it Plasma Sources Sci. Technol.} {\bf 27} 054001


\bibitem{he-cs} Cross sections extracted from program MAGBOLTZ, version 7.1 June 2004, http://www.lxcat.laplace.univ-tlse.fr

\bibitem{n2-cs} Gordillo-Vazquez F J and Donk\'o Z 2009 {\it Plasma Sources Sci. Technol.} {\bf 18} 34021

\bibitem{siglo} SIGLO database, http://www.lxcat.net/SIGLO, retrieved 10 August 2018

\bibitem{nagy} Nagy O 2002 {\it Chem. Phys.} {\bf 286} 106

\bibitem{Itikawa2006} Itikawa Y 2006 {\it J. Phys. Chem. Ref. Data} {\bf 35} 31-53 
\bibitem{LxCat} The cross section is extracted from MAGBOLTZ, Biagi S F, version 8.9, August 2018, http://www.lxcat.net/Biagi

\bibitem{Phelps} Phelps A V 1994 {\it J. Appl. Phys.} {\bf 76} 747

\bibitem{Brok} Brok W J M, Bowden M D, van Dijk J, van der Mullen J J A M and Kroesen G M W 2005 {\it J. Appl. Phys.} {\bf 98} 13302

\bibitem{Sakiyama} Sakiyama Y and Graves D B 2006 {\it J. Phys. D: Appl. Phys.} {\bf 39} 3644

\bibitem{Phelps1955} Phelps A V 1955 {\it Phys. Rev.} {\bf 99} 1307 

\bibitem{Marriott1956} Marriott R 1957 {\it Proceedings of the Physical Society, Section A} {\bf 70} 288


\bibitem{brsheath} Brinkmann R P 2007 {\it J. Phys. D: Appl. Phys.} {\bf 102}(9) 093303

\bibitem{LibermanBook} Lieberman M A and Lichtenberg A J 2005 {\it Principles of Plasma
Discharges and Materials Processing} (New Jersey: Wiley)

\end{thebibliography}
\end{document}